\documentclass[11pt]{scrartcl}
\newcommand{\topicnumber}{10}
\newcommand{\icpigheader}{30$^\text{th}$ ICPIG, August 28$^\text{th}$ - Septemper 2$^\text{nd}$, 2011, Belfast, UK}
\usepackage{cite}
\usepackage{amsmath,amssymb}
\usepackage{ulem}
\usepackage{graphicx,xcolor}
\xdefinecolor{headercolor}{RGB}{128,128,128}
\usepackage[small]{caption2}
\usepackage[paper=a4paper,left=20mm,right=20mm,top=25mm,bottom=25mm]{geometry}
\newfont{\titelfont}{cmbx10 scaled 1400}
\newfont{\authorfont}{cmr10 scaled 1200}
\newfont{\institutefont}{cmsl10 scaled 1000}
\newfont{\abstractfont}{cmr10 scaled 1000}
\newfont{\captionfontICPIG}{cmr10 scaled 1000}
\newfont{\headerfont}{cmr10 scaled 800}

\newlength\abstractwidth
\newcommand{\presenter}[1]{\uline{#1}}

\makeatletter
	\renewcommand*\bib@heading{}
\makeatother
\usepackage{fancyhdr}
\begin{document}
\sloppy
\twocolumn[
\center{
{\titelfont Numerical study of secondary electron emission in a coaxial \\[0.1cm] 
radio-frequency driven plasma jet at atmospheric pressure}\\[0.5cm]

{\authorfont \presenter{T. Hemke}$^1$, J. Trieschmann$^1$, A. Wollny$^1$, 
R.P. Brinkmann$^1$, T. Mussenbrock$^1$}\\[0.5cm]

{\institutefont $^1$ Theoretische Elektrotechnik, Ruhr-University Bochum,
44801 Bochum, Germany}
}

\center{\parbox[b]{\abstractwidth}{\abstractfont
In this work we investigate a numerical model of a coaxial RF-driven
plasma jet operated at atmospheric pressure. Due to the cylindrical
symmetry an adequate 2-D representation of the otherwise 3-dimensional
structure is used. A helium-oxygen chemistry reaction scheme is applied. 
We study the effect of secondary electrons emitted at
the inner electrode as well as the inserted dielectric tube and discuss
their impact on the discharge behavior. We conclude that a proper
choice of materials can improve the desired mode of operation of
such plasma jets in terms of materials and surface processing.
}\vspace*{0.5cm}}]

\begin{figure}[t]
\center
\includegraphics[clip,width=0.80\linewidth]{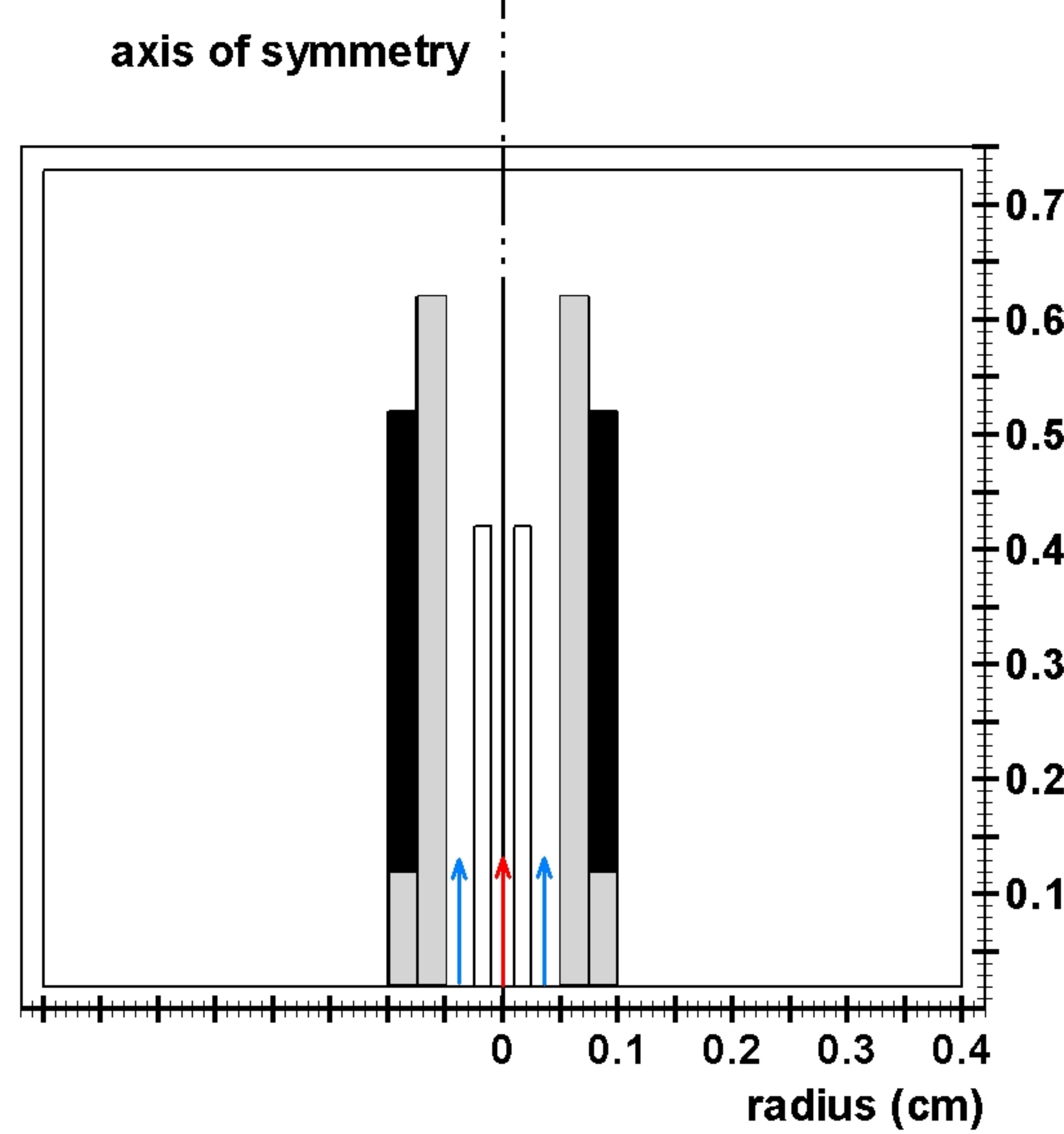}
\caption{Schematic sketch and simulation domain of the coaxial jet. 
The inner capillary (white) is electrically grounded while the 
the outer aluminum tube (black) is driven by an RF voltage. 
The ceramic tube (gray) consists of aluminum oxide.}
\label{fig:geometry}
\end{figure}

\noindent{\bf 1. Introduction}\\\indent
Microplasmas operated at atmospheric pressure have recently gained high attention \cite{Becker2010}.
A particularly popular class of microplasmas are non-equilibrium 
radio-frequency driven plasma jets, 
originally proposed by Selwyn and co-workers in 1998 \cite{Selwyn1998}. 
Many groups have investigated their prospect for surface modification;
the studied processes include etching of tungsten, deposition and 
etching of silicon oxide, and cleaning of thermolabile surfaces from contaminants.
In order to realize a jet concept for coating applications, 
e.g. homogeneous thin film deposition from C$_2$H$_2$, 
von Keudell and co-workers developed a micro-scaled axis-symmetric 
radio-frequency driven plasma jet operated at atmospheric pressure \cite{Benedikt2006}.
The schematics of their device is depicted in Fig. \ref{fig:geometry}. 
A stainless steel capillary is inserted into a ceramic tube leaving an annular 
space of 0.25 mm between the outer radius of the capillary and the inner 
radius of the ceramic tube. 
The ceramic tube is finally surrounded by an aluminum tube. 
This outer tube is now driven through a matching network by an 
RF voltage at 13.56 MHz. In contrast the inner capillary 
acts as the grounded counter electrode. 
A feed gas (helium or argon) is guided in the capillary (red arrow) and the 
annular space between the capillary and the ceramic tube (blue arrows), 
while a small amount of a reactive gas is additionally injected through the capillary.
Besides proving the successful operation of this device as depositing tool 
the coaxial jet is investigated experimentally by 
means of phase resolved optical emission spectroscopy (PROES). 
Four different modes of the discharge are distinguished in \cite{Benedikt2010}.

In this work we study the coaxial jet by means of a two-dimensional numerical 
simulation. We take advantage of the axis-symmetric geometry and resolve 
the radial and axial directions of the jet. In particular, we focus on the role 
of the secondary electrons emitted at the ceramic tube and the 
grounded capillary in order to characterize the $\gamma$-modes of the 
discharge.\\

\noindent{\bf 2. Simulation}\\
\noindent{\bf 2.1. Description of the numerical code}\\\indent
The simulations of this study are carried out with the fluid dynamics
code {\it nonPDPSIM} designed and realized by Kushner and co-workers.
For a detailed description of the code see publication \cite{Babaeva2007} 
and the citations therein. Here, we just briefly
discuss the implemented equations and the underlying physics.

The code {\it nonPDPSIM} simulates the dynamics of weakly ionized
plasmas in the regime of medium to high pressure. It takes into account
the physics and chemistry of charged particles -- electrons with mass
$m_{\rm e}$ and charge $-e$, ions with mass $m_j$ and charge $q_j$ --
and of the excited as well as the ground state neutrals (mass $m_j$). 
For all species $j$, the continuity equations (particle balances)
are simultaneously solved, where $\vec{\Gamma}_j$ is the particle
flux density and $S_j$ is the source and loss term, respectively: 
\begin{gather}
\label{eq:j_continuity}
\frac{\partial n_j}{\partial t} = -\nabla\!\cdot\!\vec{\Gamma}_j + S_j.
\end{gather}
The fluxes are calculated from the momentum balances in the
drift-diffusion approximation evaluated in the local center-of-mass
system. $D_j$ and $\mu_j$ are the diffusion constant and the mobility
(if applicable) of species $j$. Further, $\vec E$ is the electrical field,
and $\vec v$ is the mass-averaged advective velocity of the medium: 
\begin{gather}
\label{eq:j_flux}
\vec \Gamma_j = n_j \vec{v} - D_j \nabla n_j + \frac{q_j}{|q_j|}
\mu_j n_j \vec E.
\end{gather}
For the electron fluid, additionally an energy balance equation is
solved taking into account Ohmic heating and the energy losses
due to elastic and inelastic interaction with the neutrals and ions
as well as heat conduction, 
\begin{align}
\label{eq:e_energy}
\begin{split}
\frac{\partial}{\partial t} \left(\frac{3}{2}n_{\rm e} T_{\rm e}\right) &= 
\vec{j}\! \cdot\! \vec{E} \, \hdots \\ 
- \nabla \cdot \left( -\kappa_{\rm e} 
\nabla T_{\rm e} + \frac{5}{2} T_{\rm e} \vec{\Gamma}_{\rm e} \right) 
& - n_{\rm e} \sum_i \Delta \epsilon_i k_i n_i.
\end{split}
\end{align}
To capture the non-Maxwellian behavior of the electrons, all
electronic transport coefficients (the mobility $\mu_{\rm e}$,
the diffusion constant $D_{\rm e}$, and the thermal conductivity
$\kappa_{\rm e}$) as well as the electronic rate coefficients in
eqs. (\ref{eq:j_continuity}) and (\ref{eq:j_flux}) are calculated
by the local mean energy method: A zero-dimensional Boltzmann equation for
the electron energy distribution $f_{\rm}(\epsilon)$ and the transport
and rate coefficients is solved for the locally applicable gas
composition and various values of the electrical field. The tabulated
data  are then consulted in dependence of the fluid dynamically
calculated electron temperature $T_{\rm e}$. 

The plasma equations are coupled to a modified version of the
compressible Navier-Stokes equations which are solved for the gas density
$\rho$, the mean velocity $\vec v$, and the gas temperature $T$.
The contributions to the
energy equation from Joule heating include only ion contributions;
the heat transfer from the electrons is included as a collisional
change in the enthalpy. 
The scalar pressure $p$ is given by the ideal gas law. 

Finally, the potential $\Phi$ is calculated from Poisson's equation.
(The code works in the electrostatic approximation such that
$\vec E=-\nabla\Phi$.) The charge density on its right hand side stems from
the charged particles in the plasma domain and from the bound charges
$\rho_s$ at the surfaces. The coefficient $\varepsilon=\varepsilon_0
\varepsilon_{\rm r}$ represents the permittivity of the medium:
\begin{gather}
-\nabla \cdot \varepsilon \nabla \Phi = \sum_j q_j n_j + \rho_{\rm s}.
\end{gather}
The surface charges are governed by a separate balance equation,
where $\sigma$ is the conductivity of the solid materials and the
subscript $s$ indicates evaluation on the surface:
\begin{gather}
\frac{\partial \rho_{\rm s}}{\partial t} = \left[ \sum_j q_j 
(-\nabla \cdot \vec{\Gamma}_j + S_j) - 
\nabla \cdot \left( \sigma (-\nabla \Phi) \right) \right]_{\rm s}.
\end{gather}

The dynamical equations are complemented by an appropriate set of
boundary conditions. Electrically, the walls are either powered or
grounded. With respect to the particle flow, they are either solid,
or represent inlets or outlets: The  flow is specified to a given
flux, while the outlet flow is adjusted to maintain the pressure.
Finally, it is worth mentioning that the actual implementation of
the equations poses some difficulties due to the vast differences
in the time scales of the dynamics of the plasma and the neutrals.
These difficulties are overcome by the methods of time-slicing
and subcycling.\\

\noindent{\bf 2.2. Details of the simulation case}\\\indent
The described model is employed to simulate the coaxial jet described before. 
Fig. \ref{fig:geometry} provides (besides the schematic sketch of the jet) 
the simulation domain.
We scaled down the length of the original experimental jet configuration 
in order to reduce the number of grid nodes but 
preserved the original radial dimensions. 
In our case, the capillary is 4.25 mm long, the grounded electrode is 5.25 mm 
and the ceramic tube is 6.25 mm long, respectively.
The capillary has an inner and outer radius of 0.1 mm and 0.25 mm, 
the inner radius of the ceramic tube is 0.5 mm.
The simulation resolves the two cylindrical coordinates R and z. 
Due to the azimuthal symmetry of the simulation "three-dimensional physics", e.g. spot attachment of the discharge at a surface, can not be resolved.

The device is operated in a pure helium atmosphere at 10$^5$ Pa 
with helium as feed gas and oxygen as the reactive gas.
The feed gas in the annular space between the capillary and the ceramic tube 
is injected with a flow rate of 3 slm, corresponding to a maximum 
advective velocity of 80 m/s. To avoid turbulences of the gas flow 
at the tip of the capillary the flow rate of the gas injected in the capillary 
is set to maintain a similar advective velocity: Helium with an admixture 
of 1~\% oxygen as reactive gas is guided in the capillary with a 
flow rate of 170 sccm. The outlet is controlled to maintain a constant pressure.

We apply a helium-oxygen chemistry reaction scheme which is described in detail 
in \cite{Babaeva2007A}.

To study the role and influence of the secondary electrons on both surfaces 
-- the grounded capillary and the ceramic tube -- we vary the secondary electron emission (SEE) 
coefficient $\gamma$. Additionally, we weight $\gamma$ for each ion species 
to account for the different probabilities of emission of secondary electrons depending 
on the ion species hitting the surface.\\

\noindent{\bf 3. Simulation results}\\
We ignite the discharge numerically by seed electrons in the annular 
space between the capillary and the ceramic tube. 
The main ionization process during the numerical convergence 
changes from volume to secondary electron dominated ionization. 
This characteristic is found in the corresponding experiment 
by increasing the applied voltage.
In the converged state the discharge runs in the $\gamma$-mode as 
observed by the experiment.

Following, two distinct discharge regimes are discussed with simulation
results. First, the SEE coefficient 
for the ceramic tube is set to $\gamma_{\rm{cer}}=0$, 
while in the second setting it is set to 
$\gamma_{\rm{cer}}=0.1$.
The SEE coefficient for the capillary is $\gamma_{\rm{cap}}=0.1$
in both cases. Apart from that all other simulation parameters are
kept constant for both cases.

\begin{figure}[t]
\center
\includegraphics[clip,width=0.90\linewidth]{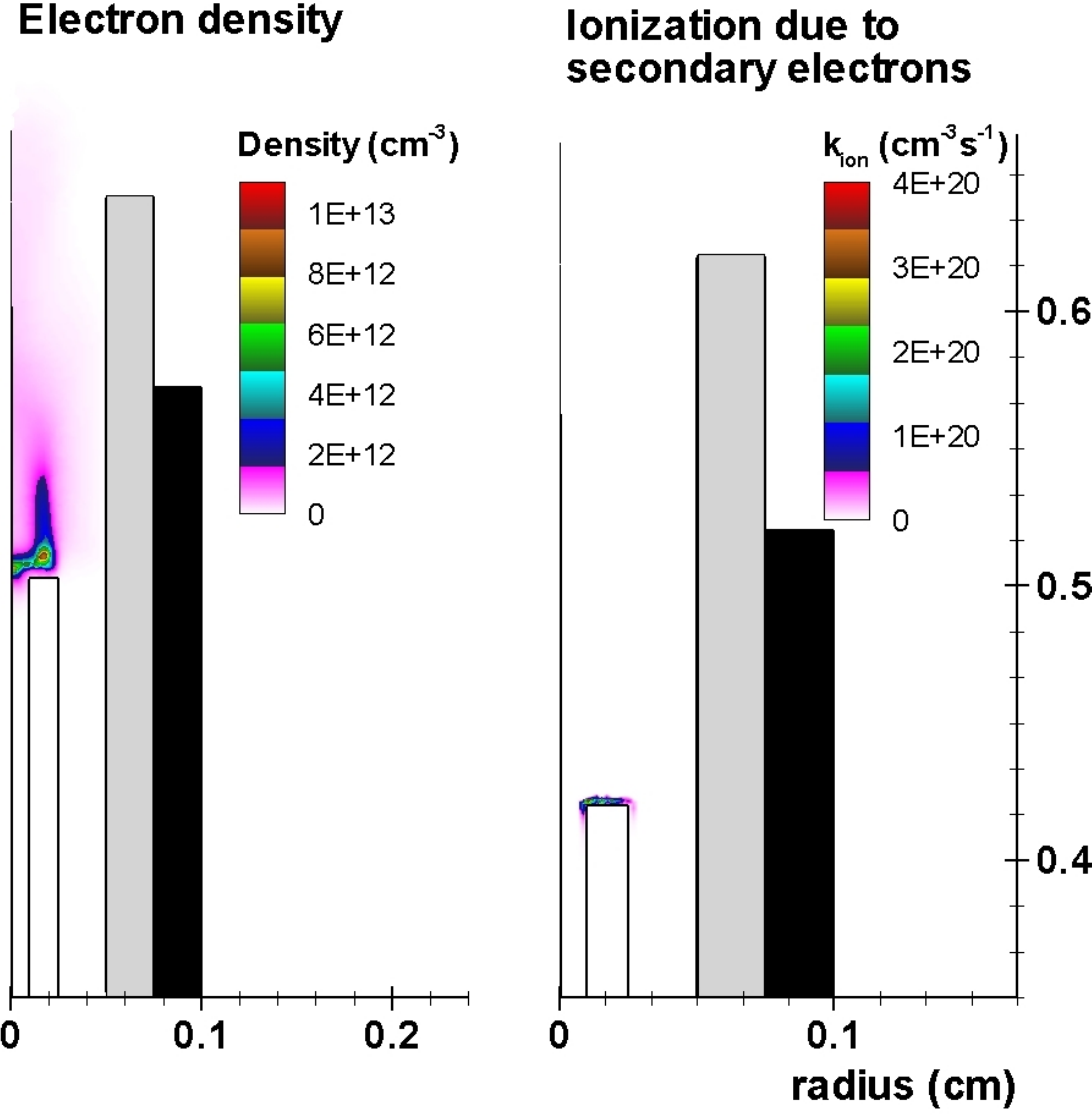}
\caption{Electron density (cm$^{-3}$) and ionization coefficient due 
to secondary electrons (cm$^{-3}$s$^{-1}$) without secondary 
electron emission at the ceramic tube.}
\label{fig:electron_dens}
\end{figure}

\begin{figure}[h!]
\center
\includegraphics[clip,width=0.90\linewidth]{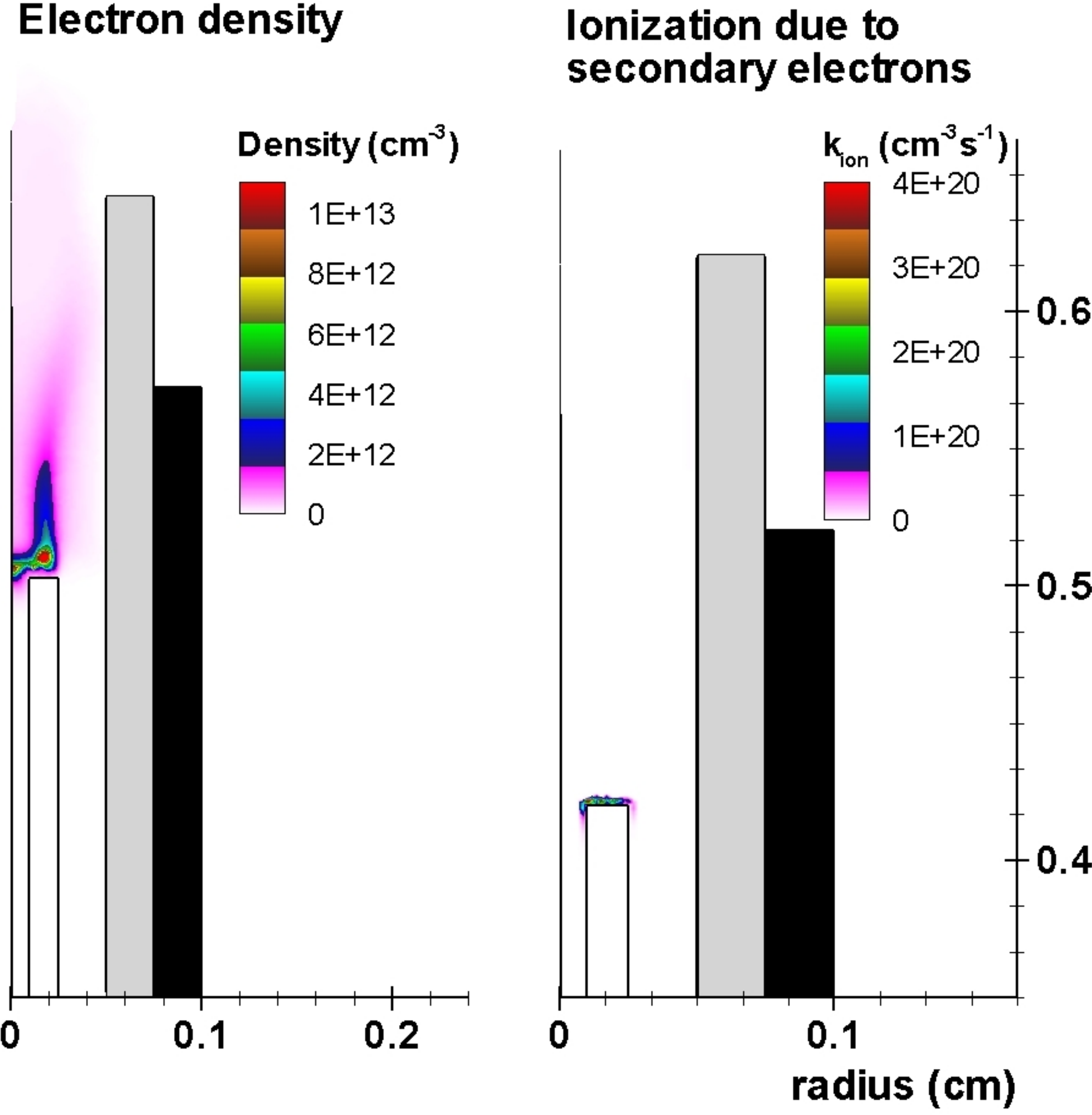}
\caption{Electron density (cm$^{-3}$) and ionization coefficient due 
to secondary electrons (cm$^{-3}$s$^{-1}$) with secondary electrons at the 
ceramic tube.}
\label{fig:electron_dens_se}
\end{figure}

\begin{figure}[t]
\center
\includegraphics[clip,width=0.90\linewidth]{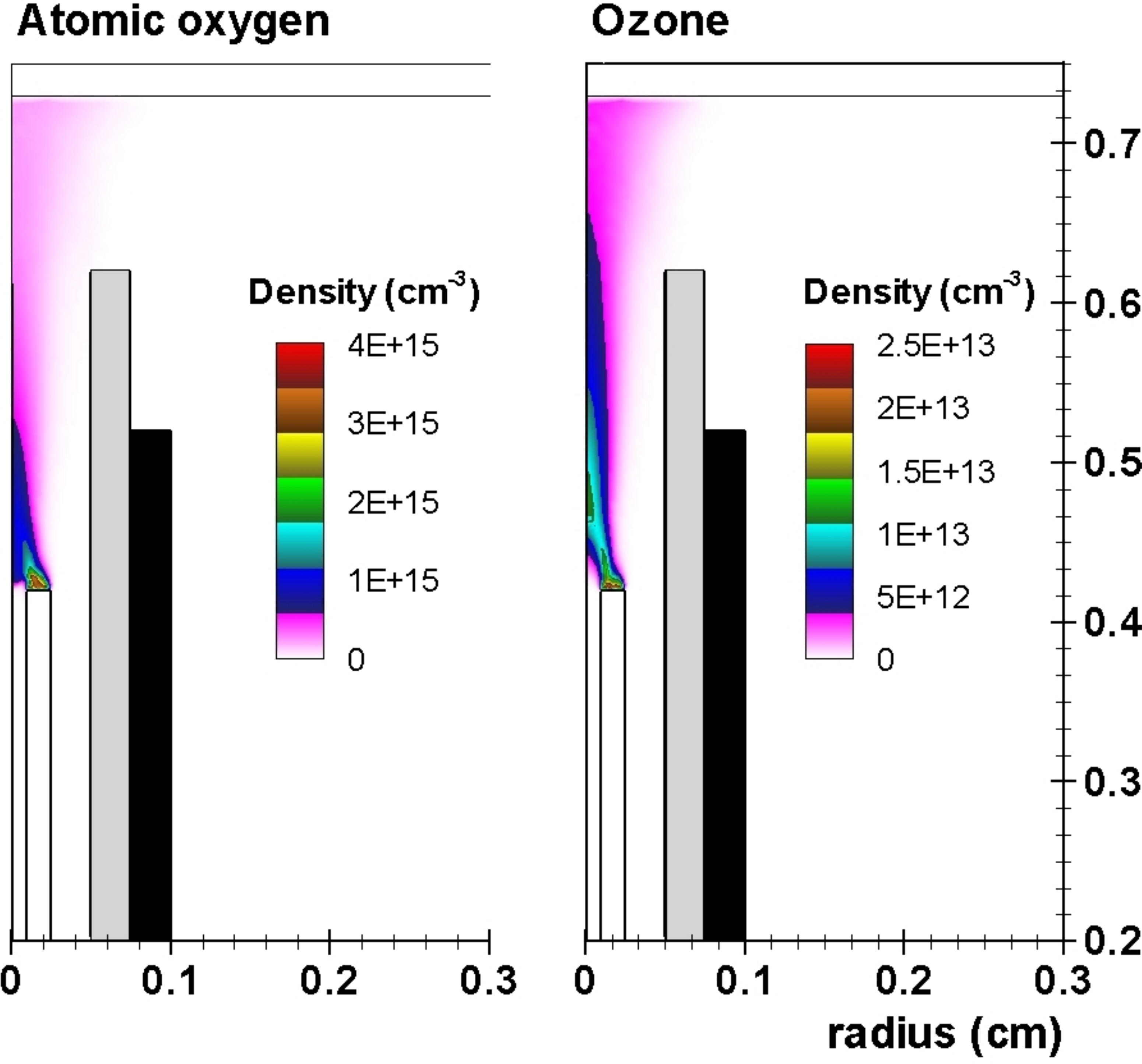}
\caption{Distribution of reactive oxygen species (cm$^{-3}$)
in the effluent of the coaxial jet without secondary 
electron emission at the ceramic tube.}
\label{fig:reac_oxygen}
\end{figure}

\begin{figure}[h!]
\center
\includegraphics[clip,width=0.90\linewidth]{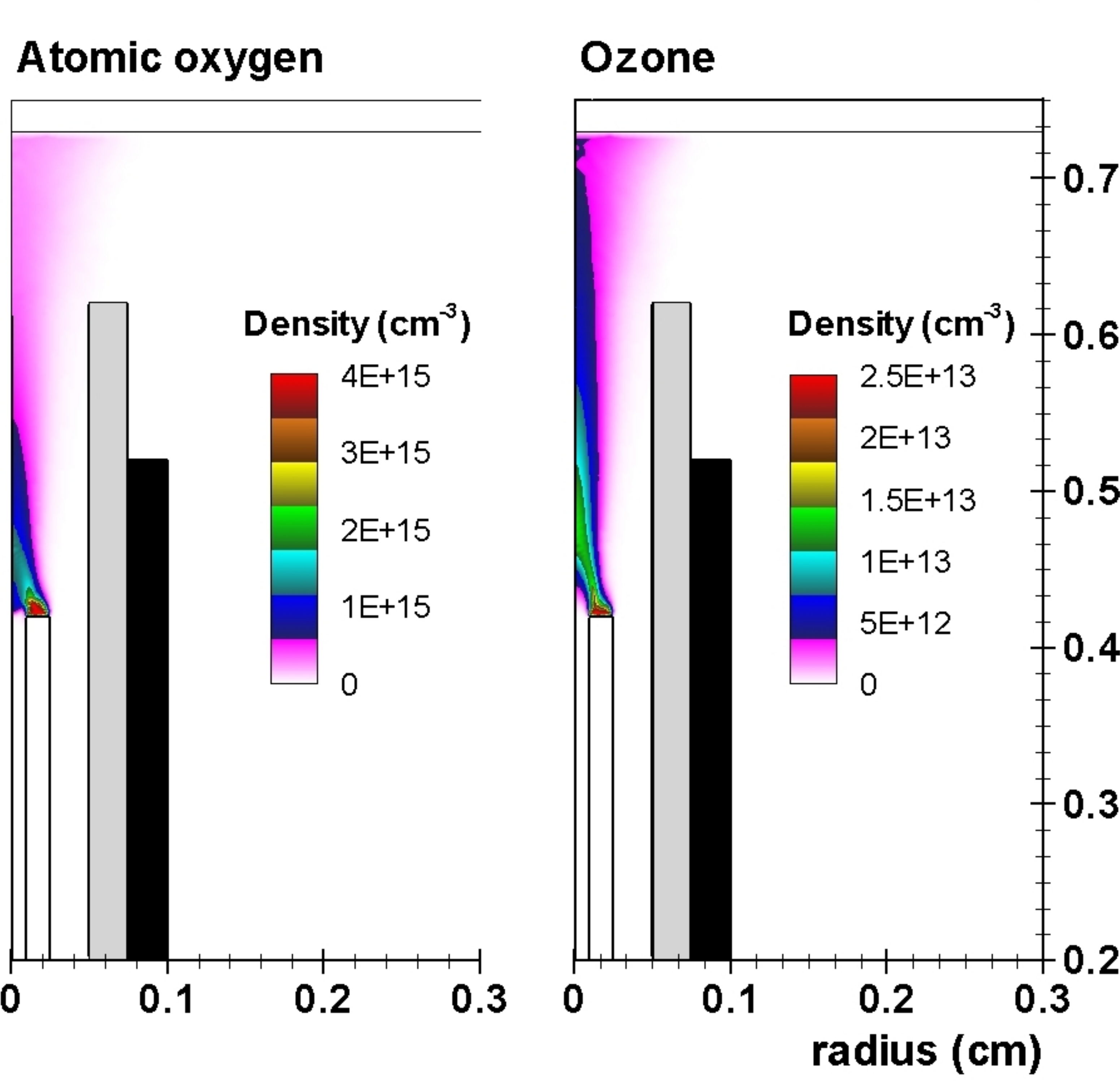}
\caption{Distribution of reactive oxygen species (cm$^{-3}$)
in the effluent of the coaxial jet with secondary electrons 
at the ceramic tube.}
\label{fig:reac_oxygen_se}
\end{figure}

In order to investigate the influence of the secondary electrons
emitted from the ceramic tube on the discharge, two aspects are
characterized. As presented first in Figs. \ref{fig:electron_dens} and
\ref{fig:electron_dens_se} the spatial profile of the electron density
as well as the ionization due to secondary electrons is analyzed.
This comparison is then followed by a comparison (compare Figs.
\ref{fig:reac_oxygen} and \ref{fig:reac_oxygen_se}) of the dominating
chemically active species, namely atomic oxygen and ozone.

In both scenarios the influence of the SEE at the ceramic tube is
clearly evident. E.g., when comparing the electron density distributions
a shift towards the ceramic tube is clearly visible with SEE at
the ceramic tube. This strongly indicates that these emitted electrons
do play a significant role in the spatial distribution of the plasma
discharge, bending it towards the surrounding ceramic tube.

With respect to reactive species the secondary electrons emitted from
the ceramic tube again play an important role. 
As can be seen in the spatial profile of ozone (compare Fig. \ref{fig:reac_oxygen_se}), 
which expands significantly further into the effluent of the
plasma jet.

These results suggest that the SEE from materials adjacent to the plasma
in a coaxial setup has a significant influence on the spatial
profile of the plasma discharge. In particular a proper choice of these
materials can alter the discharge behavior with respect to the reactive
species which finally determine the applicability in terms of materials
and surface processing.\\

\noindent{\bf 4. Conclusions}\\
In this work we investigate a numerical model of a plasma
jet in coaxial configuration operated at atmospheric pressure.
We show that the emission of secondary electrons at the ceramic
tube adjacent to the plasma discharge has substantial
impact on the spatial distribution of the plasma as well as the
reactive species, e.g. ozone. Further we conclude that a proper
choice of materials can improve the desired mode of operation of
such plasma jets in terms of materials and surface processing.\\

\noindent{\bf 5. Acknowledgment}\\
The authors gratefully acknowledge fruitful discussions with
Prof. M.J. Kushner and Dr. N.Y. Babaeva from the University of
Michigan at Ann Arbor. Financial support by the Deutsche
Forschungsgemeinschaft in the frame of Research Group 1123
{\it Physics of Microplasmas} is also acknowledged.\\


\noindent{\bf References}
\begin{thebibliography}{x}
\bibitem{Becker2010} K.H. Becker, H. Kersten, J. Hopwood, J.L. Lopez, 
Eur. Phys. J. D {\bf 60} (2010) 437439
\bibitem{Selwyn1998} S.E. Babayan, J.Y. Jeong, V.J. Tu, J.Park, G.S. Selwyn, R.F. Hicks, 
Plasma Sources Sci. Technol. {\bf 7} (1998) 286
\bibitem{Benedikt2006} J. Benedikt, K. Focke, A. Yanguas-Gil, A. von Keudell, 
Appl. Phys. Lett. {\bf 92} (2006) 251504
\bibitem{Benedikt2010} J. Benedikt, S. Hofmann, N. Knake, H. B\"ottner, 
R. Reuter, A. von Keudell, V. Schulz-von der Gathen,
Eur. Phys. J. D {\bf 60} (2010) 539
\bibitem{Babaeva2007} N.Y. Babaeva, R.A. Arakoni, M.J. Kushner,
J. Appl. Phys. {\bf 101} (2007) 123306
\bibitem{Babaeva2007A} N.Y. Babaeva, M.J. Kushner,
J. Appl. Phys. {\bf 101} (2007) 113307
\end{thebibliography}
\end{document}